\documentclass[conference]{IEEEtran}
\IEEEoverridecommandlockouts
\usepackage{cite}
\usepackage{amsmath,amssymb,amsfonts}
\usepackage{algorithmic}
\usepackage{graphicx}
\usepackage{float}
\usepackage{graphicx}
\usepackage{textcomp}
\usepackage{tabularx}
\usepackage{graphicx}
\usepackage{listings}
\usepackage{color}

\definecolor{codegray}{rgb}{0.5,0.5,0.5}

\lstdefinestyle{mystyle}{
    backgroundcolor=\color{white},   
    commentstyle=\color{codegray},
    keywordstyle=\color{blue},
    numberstyle=\tiny\color{codegray},
    stringstyle=\color{red},
    basicstyle=\ttfamily\footnotesize,
    breakatwhitespace=false,         
    breaklines=true,                 
    captionpos=b,                    
    keepspaces=true,                 
    numbers=left,                    
    numbersep=5pt,                  
    showspaces=false,                
    showstringspaces=false,
    showtabs=false,                  
    tabsize=2
}

\usepackage{xcolor}
\def\BibTeX{{\rm B\kern-.05em{\sc i\kern-.025em b}\kern-.08em
    T\kern-.1667em\lower.7ex\hbox{E}\kern-.125emX}}
\begin{document}

\title{Analyzing the Impact of Climate Change with major emphasis on pollution: A Comparative Study of ML and Statistical Models in Time Series Data}

\author{
\IEEEauthorblockN{1\textsuperscript{st} Anurag Mishra, 2\textsuperscript{nd} Ronen Gold, 3\textsuperscript{rd} Sanjeev Vijayakumar}
\IEEEauthorblockA{\textit{Golisano College of Computing and Information Sciences, Rochester Institute of Technology}\\
Rochester, NY, USA \\
\{am2552, rgg2706, sv8958\}@rit.edu}
}


\maketitle
\begin{abstract}
Industrial operations have grown exponentially over the last century, driving advancements in energy utilization through vehicles and machinery.This growth has significant environmental implications, necessitating the use of sophisticated technology to monitor and analyze climate data.The surge in industrial activities presents a complex challenge in forecasting its diverse environmental impacts, which vary greatly across different regions.Aim to understand these dynamics more deeply to predict and mitigate the environmental impacts of industrial activities.

\end{abstract}





\section{Introduction}
In an era marked by significant industrial expansion, the consequent rise in environmental pollutants has precipitated a critical need for advanced methodologies to analyze and forecast the environmental impacts of these activities. This research paper presents a comparative analysis of machine learning and statistical models to study the effects of pollution on climate change, utilizing time series data. The focus is on understanding how different modeling approaches—specifically Long Short-Term Memory networks (LSTMs), Spiking Neural Networks (SNNs), and Autoregressive Integrated Moving Average (ARIMA) models—can be employed to predict future climate conditions in response to varying pollution levels. Our study leverages two key datasets: the Earth Surface Temperature Data, which provides detailed global temperature records, and the Global CO2 Dataset, which offers geolocated information on CO2 emissions. By integrating these datasets, we aim to offer insights into the direct relationship between CO2 emissions and temperature changes, thereby supporting the development of targeted environmental policies to mitigate the adverse effects of industrial pollution.

\subsection{Problem Statement}
Climate change is not a problem that humanity is not aware about. But it is important to stablish how significant it is to know the severity of this problem so that most of the policies that people want to create our aim towards solving this crisis at hand . Over the past century, there has been a notable increase in the growth of industrial operations. With the use of sophisticated technology, we can now more efficiently harness energy for the operation of industrial machinery and automobiles in our daily life. But there are environmental consequences to this increase in industrial activity that need to be carefully considered. Our team wants to use the most recent machine-learning techniques to evaluate available climate data and develop a predictive model that can identify the direction of climate change in relation to pollution data particular to a certain region. 

Our project will find trends in the data to forecast future changes in climate and give our team the tools to analyse possible environmental effects of industrial operations in various locations.Hence resulting in a comparative study which will compare different aspects of machine learning and how machine learning can emphasise the severity of the issue of climate crisis, this comparative study will also give us a deep insight into how different models use different techniques to predict and forecast the quantities and how they deal with time series data. Time series has been challenging in the days of AI revolution because the current state of the art transformer models struggle with it and hence we as a community haven't found a really good model that handles time series data and gives us good predictions over a period of time this project aims to use this opportunity. With du dive into the ability of multiple models like GRUs and LSTMs and even biologically inspired spike neural networks as well as statistical machine learning models like the ARIMA(auto regressive integrated moving average) and the SARIMA(seasonal autoregressive integrated moving average) models and compare them and how they tackle and what are the ability they possess in handling time series data some models are highly computational while others are light on the computation but might result in a competitive result this curiosity to compare these models and also at the same time stablish the core of the climate crisis with the help of data set in regards to pollution as well we tend to move towards a project that is not just a comparative study but concrete proof of how real the problem of climate crisis is.

\subsection{Hypotheses}

\textbf{H1:} Approaching the problem of global warming by forecasting global average temperature forecasts until the year 2100 using Statistical Methods, Biologically-inspired models, and Transformers to enhance or debate climate predictions and provide hotspot regions of impact.

\textbf{H2:} Aiming to quantify the impact of pollutants using the dataset \cite{dataset-1} in the escalating trend of global warming, thereby streamlining the strategic focus for impactful measures to mitigate global warming effects.

\textbf{H3:} Through an exploration of diverse approaches to the same problem, a comparison and contrast of the algorithms presented is aimed to be conducted.

\section{Literature Survey}

One of the major contribution of the Paper \cite{san1} is the robust review on their performances whilst elaborating on many intricate aspects of Spike Neural Networks. Moreover, since SNNs are based on biological neurons, the paper also provides a comprehensive overview of the exchange of signals between two biological neurons, in addition to summarizing their detailed orientation with respect to potential differences. Specifically, they focus on Spiking Neural Network Models, Spike Neural Network Learning Mechanisms, and Spike Encodings. The equations governing the implemented model, the Leaky Integrate and Fire (LIF) are depicted in equations \ref{eq:1.1} and \ref{eq:1.2}. A more specific approach of Spike Neural Networks in its time-series forecasting task has been depicted in Paper \cite{san2}. This study focuses on using one-step ahead forecasting problems by combining a PulseWidth Modulation based encoding-decoding algorithm integrated with a Surrogate Gradient Method.They are known to induce long latency in addition to an ultra-low power consumption advantage because of using a SNN, thereby ensuring consistent spikes necessary for efficient forecasts. 

\begin{equation}
C_m \frac{dV_m(t)}{dt} = I_{ion}(t) + I_{syn}(t)
\label{eq:1.1}
\end{equation}

\begin{equation}
I_{ion}(t) = G_K n^4 (V_m - E_K) + G_{Na} m^3 h (V_m - E_{Na}) + G_L (V_m - E_L)
\label{eq:1.2}
\end{equation}

Divulging from the central idea of spike neural networks, Paper \cite{san4} proposes a framework consists of a hybrid module with each module consisting of a position-aware dilated CNN and a hop/non-hop auto-regression component combined with residual learning. Equation \ref{eq:4.1} is used as an embedding layer where $W_p$ is a learnable parameter matrix. Further, equation \ref{eq:4.2} represents the combination of inputs using residual learning. The paper \cite{anurag-1} lies in the foundation of how we are implementing our statistical modelling and the importance of basic moving average component and auto correction function, used in the deployment of autoregressive integrated moving average algorithms. Further, with the help of seasonal autoregressive integrated moving average this paper was able to forecast and predict that there is a significant seasonal pattern of rainfall.
\begin{equation}
p_i = \psi(e_i) = e_i^T W p
\label{eq:4.1}
\end{equation}

\begin{equation}
x_{i+1} = \text{ReLU}(x_i + \phi(x_i))
\label{eq:4.2}
\end{equation}

The paper \cite{anurag-3} proposes a novel approach in the realm of prediction performance in multivariate time series task with the fundamental focus on temporal change in the information. Though this paper might be focused more on LSTM for long short term memory network but it helps in analyzing how we need to handle the tiny or small amount of discrete changes in our data. It also brings forth an objective function modification where mean absolute error and mean squared error losses are integrated within the object function to evaluate the amplitude of errors. Further, the paper \cite{anurag-4} specializes in environmental forecasting. The study covers a broad spectrum and multiple 16 environment variables across the frequencies and hence evaluate the performance of different models on the environmental data set it uses one of our own data set and hence is what makes this paper really important for our literature survey.

A novel proposal by the paper \cite{anurag-7} introduces the ClimateLearn a Benchmarking Machine Learning for Weather and Climate Modeling. The library aims to bridge the gap and producibility and standardization of deep learning models for providing quantitative and qualitative analysis of climate and weather modeling task. This library helps us not only have an evaluation criteria already fixed inside the library but also makes us work really really easy since it helps us implement visualization and quantification of data models that we will be building without writing the basic evaluation criteria functions by ourselves. Finally, the paper concludes by addressing the need for standardized and reproducible framework for applying machine learning and deep learning climate science models for climate modelling and hence it produces us with an open source library that runs on pytorch and has ample amount of evaluation criteria is within itself as well as training setups and data set handling features for climate modelling which will be really helpful for our project.

Paper \cite{ronen - 2} evaluates the accuracy of CNNs and ANNs for weather prediction and compares which model is better for accurate predictions. CNNs are classifiers based on their ability to handle non-stationary and unstable datasets by adjusting model parameters. ANNs are highlighted for their error minimization capabilities and ease of instantiation in weather prediction techniques. The comparative analysis from CNNs to ANNs highlights which is better regarding climate change data classification. In one instance, CNN's ability to classify weather-dependent recognition on images showed a high accuracy of 82.2 percent, which exceeded traditional ML methods. Deviating off-course, and delving in to theoritical physics model, the paper \cite{ronen - 3} incorporates physics laws directly into machine learning for weather and climate modeling. The algorithms cover GANs, LSTMs, and CNNS with custom-design loss functions, physics regularizations, and parameter-specific model architectures that represent physics constraints.

When coming to environmental models in particular, the paper \cite{ronen - 7} complies multiple deep-learning techniques to identify leaf stressing in different plants with CNN architectures. The research evaluates the techniques based on stress type, dataset size, training, test sets, and the model used. The purpose is to highlight the usefulness of machine learning in agriculture and provide a benchmark for readers to understand how this technology could be applied to improve our environment further. Papers like these enables this study to be led in this direction to allow for an efficient climate predictor model to suggest preventive measures in advance.

\section{Method}

\subsection{Datasets}
\textbf{Dataset 1: Climate Change: Earth Surface Temperature Data }
\begin{itemize}
    \item The dataset encompasses a total of 1.6 billion samples. Considering its archaic nature, it will really benefit in capturing long-term and short-term trends. Further, it also addresses the historical complexities such as the uncertainty in values captured. Moreover, basing on the various features available, the dataset has been classified in to the following:
    \begin{itemize}
        \item \texttt{GlobalTemperatures.csv} - Global Land and Ocean-and-Land Temperatures
        \item \texttt{GlobalLandTemperaturesByCountry.csv} - Global Average Land Temperature by Country
        \item \texttt{GlobalLandTemperaturesByState.csv} - Global Average Land Temperature by State
        \item \texttt{GlobalLandTemperaturesByMajorCity.csv} - Global Land Temperatures By Major City
        \item \texttt{GlobalLandTemperaturesByCity.csv} - Global Land Temperatures By City
    \end{itemize}
    \item The important features include from the primary file \\ (\texttt{GlobalTemperature.csv}) include:
    \begin{itemize}
        \item \textbf{Date(YYYY-MM-DD)}: We aim in pre-processing with Datetime Feature engineering, that involves, fabricating datasets with different frequency intervals to expand the learning range of the model. In addition to this, the dataset we plan on adding a time lag variable, that refurnishes the dataset with each sample
    \end{itemize}
\end{itemize}

\textbf{Dataset 2: Global Air Pollution Dataset}

\begin{itemize}
    \item This dataset presents geolocated information on air pollution, a critical environmental issue affecting public health.
    \item It includes data on major pollutants: Nitrogen Dioxide (NO2), Ozone (O3), and so on.
    \item The dataset features AQI values and categories, providing an overall assessment of air quality in various countries and cities...
    \item The prime features include: Country: Name of the country, City, AQI Value, AQI Category, CO AQI Value, CO AQI Category, Ozone AQI Value, Ozone AQI Category, NO2 AQI Value, NO2 AQI Category, PM2.5 AQI Value, PM2.5 AQI Category
\end{itemize}
Our ultimate goal is to present a study by training models on these two datasets and analyze the pollution trend over the years contributing to the upcoming rise in Average Global Temperatures, caused by global warming. Based on this, upon compiling data from the 1700s until the present, we aim to come up with a factor of contribution (FOC) of pollution to global warming.\\
\sloppy
\subsection{Preprocessing}

\textbf{Sanjeev Vijayakumar}

The dataset spans time series values from the 1800s to 2014, with each sample representing a day in history and containing average values for that area. It encompasses global temperatures across countries, cities, and major cities. For training, I utilized the dataset clustered over major cities, providing ample data. Using pandas, I split it into 100 datasets based on cities, enabling sequential model training for generalization and smooth forecasting.

To process the data for spike neural networks, I employed temporal encoding that are summarized as follows:

\subsection{Spike Neural Networks}
\begin{enumerate}
    \item Normalizing data within the range [0, 1] is crucial to avoid gradient explosion in Spike Neural Networks, which output discrete spikes.
    \item A threshold of 0.5 enables the neuron voltage to generate spikes.
    \item Initializing the \texttt{TIME\_STEPS} variable, temporal encoding maps each sample into an array spanning \texttt{TIME\_STEPS}, with spikes occurring at different intervals.
    \item Encoding each sample into a time series input includes the \texttt{TIME\_WINDOW} variable to capture statistical data dependencies.
\end{enumerate}

\subsection{ARIMA}

ARIMA (AutoRegressive Integrated Moving Average) is a forecasting technique that models time series data based on differences between observations (integrated), trends (autoregressive), and noise (moving average). ARIMA Parameters (p, d, q):
\begin{enumerate}
    \item p: Autoregressive terms (lags of the series)
    \item d: Differencing order (number of times the data have had past values subtracted) 
    \item q: Moving average terms (lags of the forecast errors)
\end{enumerate}  
We need a way to integrate and test whether the data given to us is stationary or not to check that the data given to us is stationary or not 
we did the Utilized Augmented Dickey-Fuller (ADF) test to check for stationarity, this gave us the result that the data was stationary. Hence our model could be applied to it as both.ARIMA and SARIMA require the data to be stationary for it to be applied and get the predictions out of it otherwise the  statistical model will fail and will not give correct predictions. To incorporate CO2 emissions required method so that correlation could be established with temperatures and to do that. We need to go deeper and search for a tool inside the SARIMA model to incorporate the correlation and hence we found exogenous variable that we can use to incorporate the emissions by country whose city we are predicting the temperature for hence we Integrated annual CO2 emissions as an exogenous variable in the SARIMAX model.

\textbf{Ronen Gold}

The first method used to forecast temperatures is a Long Short-Term Memory (LSTM) neural network that aims to map the temporal relationships between air pollution and temperature. The data was preprocessed using different techniques for input to the LSTM model. The temperature data was labeled with the date (YYYY-MM-DD), city, country, and respective temperature collected dating back to the 1800s, while the pollution data has geolocated information on the CO2 emissions of the respective city and country.

\subsection{Methodology}

\textbf{Sanjeev Vijayakumar}

With a review of Spike Neural Networks and Biologically-inspired algorithms, it became clear that Spike Neural Networks are pivotal. These networks rely on precise data pre-processing, generating spike trains that match forecast times. For instance, our dataset spans multiple days, distributed across discrete time steps. Each sample undergoes binary encoding, producing spikes at specific time intervals, facilitating a shift to time-based encoding, also known as Temporal Encoding.

Initially, we favored Spike Neural Networks in BindsNET for their efficiency and robustness. However, the Spiking Jelly Framework in Torch offered versatility with feedforward layers. Thus, our architecture, based on the Spiking Jelly framework, includes a Flattening Layer, Feedforward input and hidden neural networks, and a Leaky Integrate and Fire (LIF) Node, depicted in Figure.

\begin{figure}[H]
  \centering
  \includegraphics[width=\linewidth]{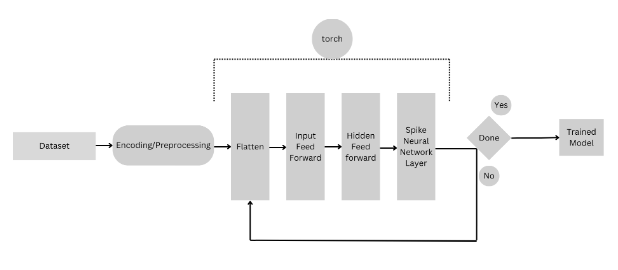}
  \caption{Spiking Neural Network Architecture. Model architecture depicted for Spike Neural Network Architecture utilized to forecast global land temperatures.}
  \label{fig:sanjeev_spike_architecture}
\end{figure}

\textbf{Anurag Mishra} \\
The AR part of the ARIMA model captures the feedback relationship between past observations and future values:

\begin{equation}
AR(p) = \phi_1 Y_{t-1} + \phi_2 Y_{t-2} + \dots + \phi_p Y_{t-p}
\end{equation}

where $\phi_1, \phi_2, \dots, \phi_p$ are the parameters of the AR part of the model, and $Y_{t-k}$ represents the value of the series at time $t-k$.

\textbf{Integrated  Term}

This refers to the number of differences required to make the series stationary:

\begin{equation}
\nabla Y_t = Y_t - Y_{t-1}
\end{equation}

For higher-order differencing:

\begin{equation}
\nabla^2 Y_t = Y_t - 2Y_{t-1} + Y_{t-2}
\end{equation}

\textbf{Moving Average (MA) Term}

This part models the dependency between an observation and a residual error from a moving average model applied to lagged observations:

\begin{equation}
MA(q) = \theta_1 \epsilon_{t-1} + \theta_2 \epsilon_{t-2} + \dots + \theta_q \epsilon_{t-q}
\end{equation}

where $\theta_1, \theta_2, \dots, \theta_q$ are the parameters of the MA part of the model, and $\epsilon_{t-k}$ is the white noise error term at time $t-k$.

\textbf{Combining the Components}

The full ARIMA model combines these components into a single equation:

\begin{equation}
Y_t' = c + \phi_1 Y_{t-1}' + \dots + \phi_p Y_{t-p}' + \theta_1 \epsilon_{t-1} + \dots + \theta_q \epsilon_{t-q} + \epsilon_t
\end{equation}

where $Y_t'$ is the differenced series, $c$ is a constant, and $\epsilon_t$ is the error term at time $t$.
Hence in Conclusion The ARIMA model's ability to model and forecast time series data by accounting for various underlying patterns makes it a powerful tool in quantitative analysis. Proper selection of its parameters, $p$, $d$, and $q$, is crucial for achieving reliable and accurate forecasts. Hence, what we did was, we took in the data from the past and associated each variable that was required for this algorithm to work and then tried on multiple different statistical models that are the versions of Arima and then we finalized that for our first prediction, ARIMA(5,1,3) give the best results for predicting the temperature to the most accuracy and precision.

\textbf{Ronen Gold}

We isolated the relevant parts of the data to train an individual model per city. This process allowed each model to be trained on its respective temperature data to provide accurate results in the specified city. The process identified errors for the Global Temperature dataset and interpolated any missing values. We resampled the data to a monthly average to reduce noise. Normalization was applied using the MinMaxScaler to scale the feature values between 0 and 1. The normalization aids in the convergence of the network but also standardizes the distribution of the input features. To improve the predictive capability of the LSTM, we structure the scaled data into overlapping windows of 12 months that serve as input to predict the subsequent month's temperature. To increase performance in the model, the final results had an overlapping window of 120 months (10 years) that the model used for forecasting. The restructuring transforms the model into a supervised learning task where the model predicts a future value based on a sequence of past values.

LSTMs can handle long-term capabilities, making them desirable models for predicting temperatures. The models were first trained on temperature data for five random cities and then on temperature and pollution data to provide forecasts for side-by-side comparison. Based on historical data, each model was used to predict 120 months into the future. The architecture consists of the LSTM layer that loops over 50 epochs, followed by an activation layer that relies on the “tanh” function. The output layer is a fully connected dense layer that outputs a single continuous value representing the forecasted temperature of the following month. The model is compiled using the Adam optimizer and Mean Squared Error. 

\subsection{Validation}

\textbf{Sanjeev Vijayakumar}\\

Our goal is to forecast temperatures over time, with validation occurring in two stages. Firstly, by calculating global average land temperatures. High values would signal concerning levels of global warming, while lower values suggest less urgency.

Secondly, we examine trends in CO2 emissions alongside global land temperature. Consistent trends may indicate a link between emissions and temperature rise, guiding potential improvements.

The dataset is initially split into train and test sets, with the first 80\% used for training and the remaining 20\% for testing. Random shuffling ensures robustness and facilitates comparison across models. Validation is based on Mean Squared Error (MSE).\\

\textbf{Anurag Mishra}\\

The ARIMA model was rigorously evaluated to determine its forecasting accuracy on our time series data. The following metrics were computed after fitting the model on the training set and predicting the values on the test set, which was separated using an 80/20 split. Even though this statistical model could have worked without the split as well, we required to split the data set because we needed to validate and see if the best model is actually ping the data set for the best results as well. The below graph shows us how the validation actually occurred for this specifi reason and why we needed to split the data for a statically model, which is not a machine learning model, we could have used all of our data for learning and could have given predictions easily. This was usually done to reduced overfitting because even statistically models, though they have an aggregating term they are under the danger of overfitting.\\

\begin{figure}[H]
  \centering
  \includegraphics[width=0.9\linewidth, height=0.7\linewidth]{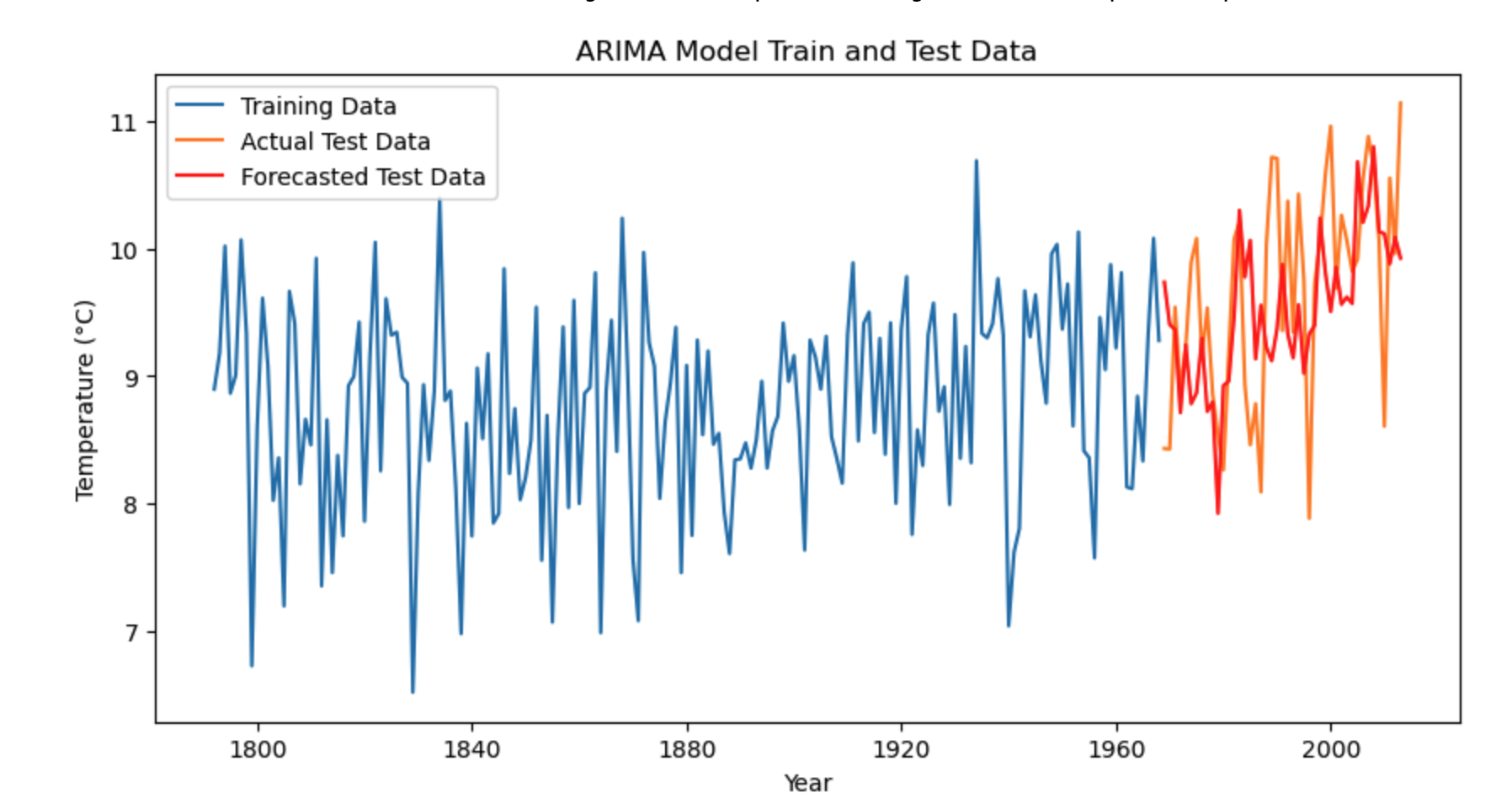}
  
  \caption{This figure represents the validation prediction of predicted versus test data for ARIMA}
  \label{fig:Validation Graph pf ARIMA }
\end{figure}
\textbf{Ronen Gold}

The validation of the LSTM occurred using a test-train split of 80-20. The test set is used to measure the level of capability the model has on unseen data. The validation loss illustrated by the figure below shows how well the model minimized its errors over epochs during the training phase. A model that rapidly decreases and stabilizes over a low value indicates that the model is efficiently training and not prone to overfitting on the data. Then, we analyze the model performance using a variety of performance metics: 

\begin{itemize}
    \item Root Mean Squared Error (RMSE): Measures the average magnitude of the errors in a set of predictions.
    \item Mean Squared Error (MSE): Squares the errors before averaging and penalizes larger errors more severely.
    \item Mean Absolute Error (MAE): Average of the absolute differences between predictions and actual observations. 
    \item R-squared (Coefficient of Determination): Indicates the percentage of the response variable variation explained by a linear model. 
    \item Explained Variance: Measures the proportion to which a mathematical model accounts for a given dataset's variation.
\end{itemize}

\section{Results}

\textbf{Sanjeev Vijayakumar}

\subsubsection{Challenges Faced}
The Spike Neural Network I utilized had one node of Leaky Integrate and Fire (LIF), which expects binary input. As a result, I had previously summarized that the data pre-processing enabled the conversion of the dataset into a binary structure suitable for the model. This only allows the model to predict binary values based on the threshold I have assigned. So in this study, for the spike neural network, I assigned a value of 0.5 to it, allowing the network to assign the binary value as 1 if it is above the defined threshold and 0 if it is lower. Due to this reason, instead of forecasting numerical values, I will only be able to allot specific bins to it. Further, with the help of the surrogate function corresponding to the LIF Node, I was able to give it an $\arctan()$ function, capable of smoothing the curve from $-\frac{\pi}{2}$ to $+\frac{\pi}{2}$. \\

\subsubsection{Depiction of results}
The model was trained over 100 epochs on 100 datasets, namely the New\_York\_data.csv, Delhi\_data.csv, Shanghai\_data.csv, and so on. The model’s architecture is depicted as follows in Figure \ref{fig:sanjeev_spike_architecture}. As shown, there are three layers: the LIF Node, a feed-forward hidden layer, and a feed-forward output layer. These layers were included to compress the input suitable for our predictions.

\begin{figure}[H]
  \centering
  \includegraphics[width=0.7\linewidth, height=0.7\linewidth]{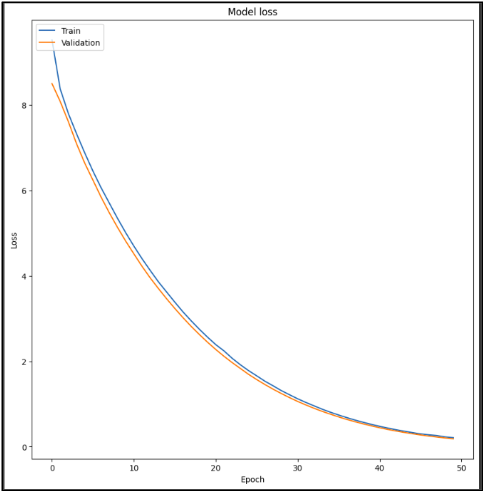}
  
  \caption{Spiking Neural Network Training/Validation LossThe figure represents the training and validation loss over 50 epochs of the training model}
  \label{fig:sanjeev_train_val_loss}
\end{figure}

Since the model was trained over all the city datasets, each spanning an average of 1500 samples, it was able to generalize well. A depiction of the training/validation curve can be observed in Figure \ref{fig:sanjeev_train_val_loss}. As can be seen, the model initially has a higher error rate but gradually smooths out due to the generalization of the model with the validation set.

Two subsets of cities were chosen to depict the results in a robust manner. Each subset comprises a list of five cities. The first subset includes a set of five chosen cities, considering their comprehensive data with minimal missing values. These cities are New York, Delhi, Shanghai, Tokyo, and Berlin. The other subset of cities was chosen randomly, consisting of Jiddah, Acapulco, Edmonton, Irkutsk, and Bac Lieu. The depiction for the first set of cities is listed as follows in Table \ref{tab:sanjeev_cities_results}. The Mean Average Error tends not to be very high; however, in the same case, the obtained values are not sufficient for long-term forecasts.

\begin{figure}[H]
  \centering
  \includegraphics[width=0.7\linewidth, height=0.7\linewidth]{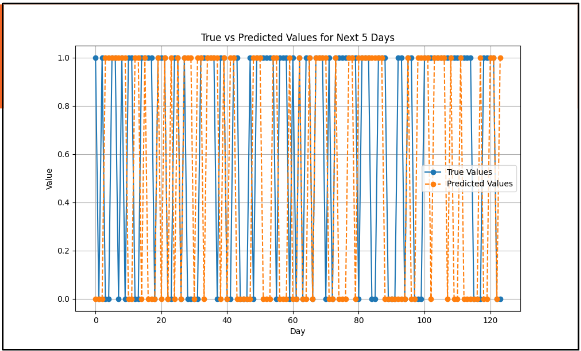}
  
  \caption{Spiking Neural Network Binary ForecastsThe figure represents the binary forecasts of the spike neural network model with a threshold of 0.5}
  \label{fig:sanjeev_binary_forecast}
\end{figure}

\begin{table}[H]
\centering
\caption{Frequency of Special Characters}
\label{tab:sanjeev_cities_results}
\small
\begin{tabular}{|l|c|}
\hline
Evaluated City & Mean Average Error (MAE) \\ \hline
New York       & 0.0812                   \\ \hline
Delhi          & 0.0712                   \\ \hline
Shanghai       & 0.0935                   \\ \hline
Tokyo          & 0.1523                   \\ \hline
Berlin         & 0.07452                  \\ \hline
\end{tabular}
\end{table}
Further, the binary forecasts have been represented in the Figure \ref{fig:sanjeev_binary_forecast}. The true values and predicted values coincide in a many samples. As the spike neural network gives a binary output, the actual continuous forecasts were converted to binary as well. Although many samples coincide, there are a lot of values classified incorrectly. In addition to this, this figure represents a scenario where the forecasts are short term (30 day time period). The results for long term forecasts were not very accurate with constant miclassifications to the binary value 0. With this analogy, this study can be dediced that the model needs more complexity with an amplitude factor encoded to it.

\begin{figure}[H]
  \centering
  \includegraphics[width=0.7\linewidth, height=0.7\linewidth]{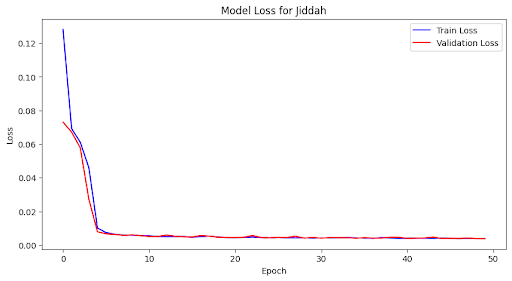}

  \caption{Validation Loss Curve for Jiddah ModelThe figure shows the validation loss curve on the validation data for the model trained on the Temperature data for Jaddih}
  \label{fig:sanjeev_continuous_forecast}
\end{figure}

Moreover, even though in the continuous forecasts figure represented by Figure \ref{fig:sanjeev_continuous_forecast}, the values seem to match in many samples. However, there are two thigs to be considered. First, this is an approximation by the surrogate function of the LIF Node. Second, the forecasts have a lot of noise associated with it, and because of this the values seems to shoot to maximum value and drop to the minimum value immediately in consequent samples. As one can deduce, a behavior such as this is not very apt when considered the temperature forecasts, thereby the results are not accurate.

\textbf{Anurag Mishra}
\subsubsection{Challenges Faced}
The challenges that I faced for this specific model is that of incorporating carbon dioxide, and its emission in the prediction, which was the most difficult part of the project since it it is designed for seasonality not for predicting for in regards to one temperature we need to do a very deep research and bring out an old way in which the status table models will be implemented with correlation to only one figure.

The ARIMA model was rigorously evaluated to determine its forecasting accuracy on our time series data. The following table presents the error metrics and model selection criteria obtained after fitting the model on the training set and predicting the values on the test set, which was separated using an 80/20 split.The results that we are getting from the statistical model of both ARIMA And SARIMA gives us an insight about the prediction, and hence there are the best results among all three models in terms of predicting the temperature into the future. This prediction has resulted and surprised all of us since we were expecting machine learning models to perform better than statistical models, but turns out that statistical models are better at predicting and learning does not solve all the problems of prediction specifically with temporal data and series data. We observe the fact that statistically models usually perform learning models, and in this case even spike neural networks. 

\textbf{Error Metrics and Model Selection Criteria of Statistical Model SARIMA/ARIMA for Berlin Model }

\begin{table}[H]
\centering
\begin{tabular}{|l|c|}
\hline
Metric & Value \\ \hline
Mean Absolute Error (MAE) & 0.068 \\ \hline
Mean Squared Error (MSE) & 0.070 \\ \hline
Root Mean Squared Error (RMSE) & 0.84 \\ \hline
Akaike Information Criterion (AIC) & 543.288 \\ \hline
Bayesian Information Criterion (BIC) & 741.945 \\ \hline
\end{tabular}\\ 

\caption{Summary of model evaluation metrics}
\label{tab:metrics}
\end{table}

Predictions indicate a potential increase in adverse climate events if current industrial growth persists without mitigative actions.
The MAE and RMSE values suggest that the model reasonably fits the historical data. Given that both are less than 1°C.

The following are the results that we get when we incorporate carbon dioxide emissions of the specific country in this case we are showing you one of the cases since there are multiple graphs of multiple countries and that does not require us to incorporate every graph possible without compromising the details of the graph in this restrain of number of pages we can use hence we are giving you one of the cities as a demonstration of how our model works. The code for this is provided and you can access all the cities you want with respect to the data set, which is very heavy and takes time to c compile.

\begin{figure}[H]
  \centering
  \includegraphics[width=0.9\linewidth, height=0.7\linewidth]{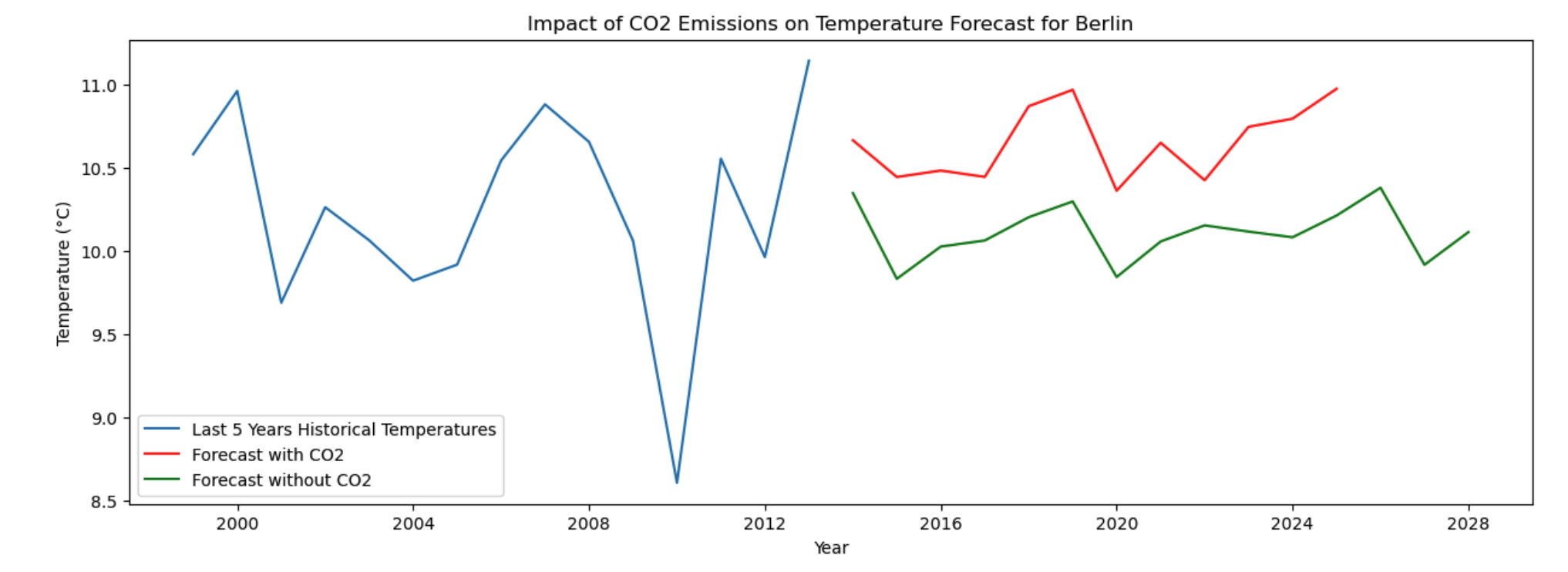}
  \caption{Predicted Co2 and Without CO2 using SARIMA(Berlin)This figure represents when we incorporate carbon dioxide and show that carbon dioxide temperatures are higher }
   \label{fig:Validation Graph pf ARIMA }
\end{figure}

The results in Figure 7 of New York where we predicted the temperature for the future while taking an account the data that was longest with us that is from the 1800s. This resulted in the one of the best predictions and hence it needs to be mentioned in the research paper, but how the statistical model predicted the very best prediction, this is another city that we are used for predicting and hence this auto aggressive moving average model predicted the result into the future. This figure shows only the last five years because that is the best way to represent it because the number of years if represented on such a small graph would resu in a cluttered graphical representation and hence we needed a model that shows us the temperatures without the carbon dioxide emission data included in it. Hence the New York predictions given below gives us an image how good the statistical models behave.

\begin{figure}[H]
  \centering
  \includegraphics[width=0.9\linewidth, height=0.7\linewidth]{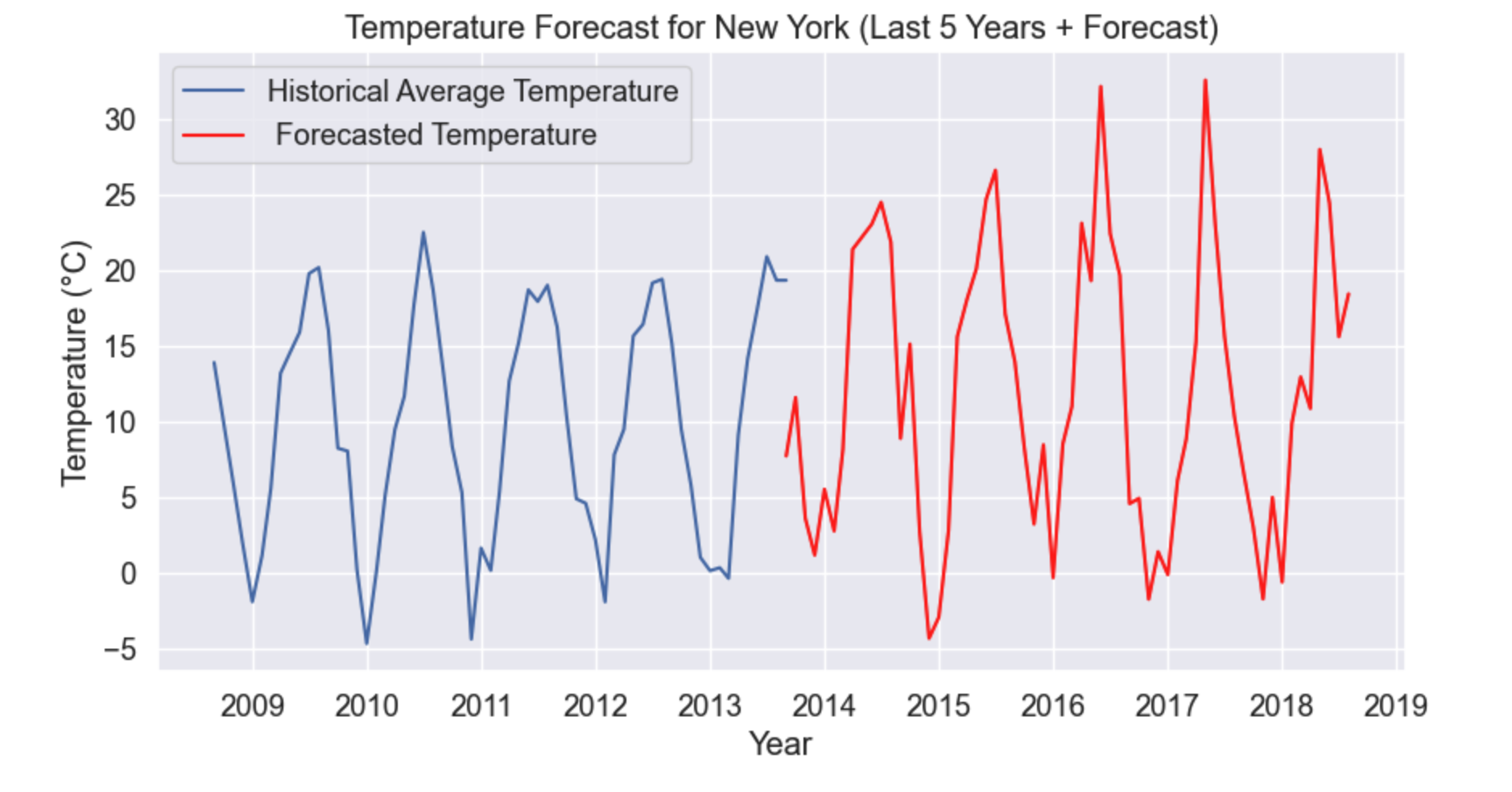}
  \caption{Prediction of future temperatures for New York using Statistical Model ARIMA.his figure represents what a simple statistical model can accurately predict the future temperature.}
  \label{fig:New York Predcition }
\end{figure}

Hence, it could be concluded that both ARIMA and SARIMA were able to incorporate and capture the time series data as required the alkaline information criteria, and the Beijing information criteria as mentioned above is the of the one best model that we used to predict the different cities required different adjustment for prediction. this is one of the best prediction and alkaline information, criterion and information critrion other matrices are reported for this best model.

\textbf{Ronen Gold}\

The results in Table 3 depict the different model's performance in each city. Most models demonstrated high predictive accuracy, evident by their high R-squared and Explained Variance values. This indicates that the models could capture the variance in temperature data quite well. Some models, like Acapulco, showed lower scores in these areas, suggesting that the model couldn’t capture the variance in the data. This could be due to a lack of data for that specific city. The RMSE and MAE values across all cities stay consistently pretty low, indicating that the model can predict forecasted temperature with some degree of significant value.

\begin{table}[H]
\centering
\caption{Performance Metrics of the LSTM Model Across Different Cities}
\label{tab:metrics}
\small
\begin{tabular}{|l|c|c|c|c|c|}
\hline
City     & RMSE    & MSE     & MAE     & R-squared & Exp Var \\ \hline
Jiddah   & 0.0691  & 0.0047  & 0.0534  & 0.9361    & 0.9483              \\ \hline
Acapulco & 0.0814  & 0.0066  & 0.0639  & 0.7836    & 0.7892              \\ \hline
Edmonton & 0.0657  & 0.0043  & 0.0484  & 0.9162    & 0.9202              \\ \hline
Irkutsk  & 0.0461  & 0.0021  & 0.0347  & 0.9683    & 0.9688              \\ \hline
Bac Lieu & 0.0568  & 0.0032  & 0.0453  & 0.8713    & 0.8849              \\ \hline
\end{tabular}
\end{table}
The model loss shown in Figure 8 for Jiddah shows two curves: one for training loss and another for validation. The decrease within the first few epochs indicates that the model adapts quickly to the patterns in the temperature data. After the decline, the model lost some stability while converging towards a low value; however, it eventually stabilized at a low value without overfitting the noise in the data. The model can capture trends in the data because of its ability to predict because of its long-term dependencies. The model was trained for five other cities, and the validation loss curves are similar across models. 

\begin{figure}[H]
  \centering
  \includegraphics[width=0.7\linewidth, height=0.7\linewidth]{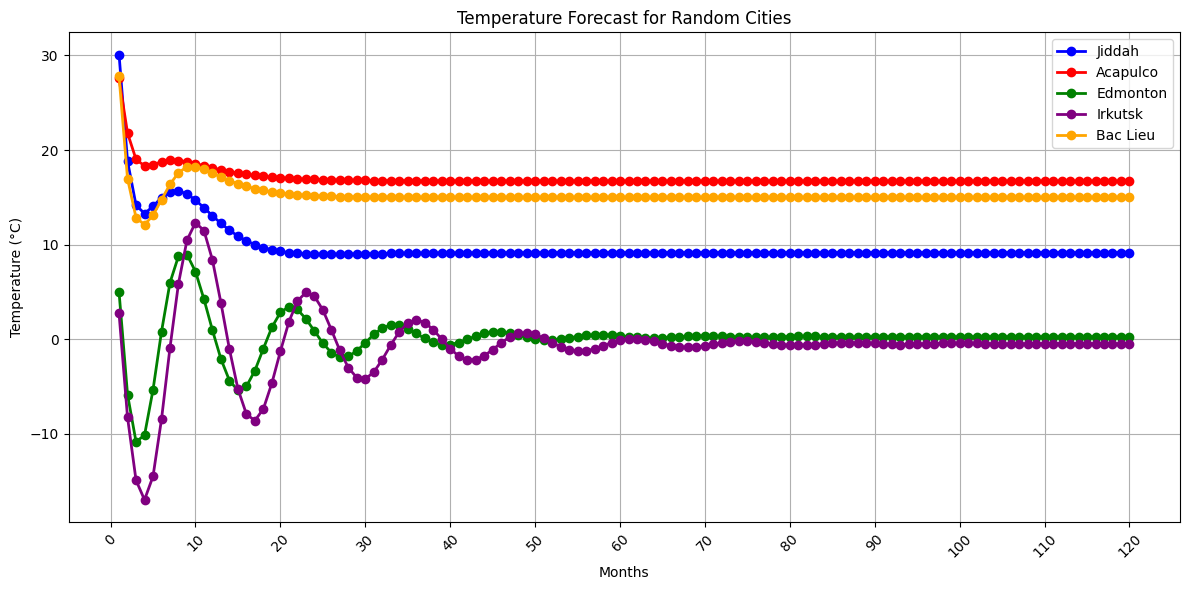}
  \caption{Temperature Forecast spanning 120 Months with LSTM}

  \label{fig:Jiddah_validation_loss}
\end{figure}

The temperature forecast in Figure 9 shows the model's predictive capability on unseen data. Each city was trained with temperature data for its specific area and used to forecast the future for 120 months. The model underfits the data quite a lot and averages out after a certain point. In some models, it averages much quicker. The best predictive capability is seen until the 12-month mark, but the predictions are very smooth from month to month.

\section{Discussion}

\subsection{Future Works}
As observed, the spike neural networks were not very efficient in computing the forecasts. Consequently, at this point, the research can explore the following possible directions:

\begin{enumerate}
    \item Implement an npc (number of parts per carrier) term, which defines an amplitude associated with each forecast value instead of just a categorical encoding.
    \item Develop an Artificial Neural Network and then convert it to a Spike Neural Network using spiking jelly, thereby allowing inherent numerical information.
\end{enumerate}
For future work, the main thing that we need to do is incorporate much more pollution data not just the carbon emission, but also ppm and other policy matrix that can help us predict the temperature even better.
\section{Conclusion}
This section presents the Mean Squared Error (MSE) performance of various models The following table summarizes the MSE results from the different predictive models utilized in our study, highlighting the performance of each in forecasting temperature data for Irkutsk.

\textbf{Mean Squared Error (MSE) Comparison}

\begin{table}[H]
\centering
\begin{tabular}{@{}lc@{}}

Model & MSE \\ 
Spiking Neural Network & 0.0257 \\
LSTM & 0.0021 \\
Statistical Model & 0.013 

\\ 
\end{tabular}
\caption{Comparison of MSE values for different predictive models}
\label{tab:mse_comparison}
\end{table}
The statistical model demonstrated the best performance among the models tested in this study, as evidenced by its MSE value of 0.013, which, while not the lowest, indicates a robust balance between sensitivity and generalization. This superior performance can be attributed to several factors:

\begin{enumerate}
    \item \textbf{Model Simplicity:} The statistical model, likely an ARIMA or similar, benefits from being highly focused on the time series data's inherent properties such as trend and seasonality, without the need for extensive data preprocessing or transformation.
    \item \textbf{Data Efficiency:} Unlike deep learning models that often require large datasets to perform well and avoid overfitting, statistical models can make more efficient use of smaller datasets and still provide accurate forecasts. This is particularly advantageous in cases like Irkutsk, where extensive historical climate data may not be abundantly available.
\end{enumerate}
Hence, had the best result, which is surprising regarding of the fact that usually it is the machine learning models that have better results this could be because statistical model had to work with lesser variable for predicting higher number of temperature points, and this is the result by both spiking neural networks, and the deep learning model resulted in overfitting and does not perform well in all the cities as well. We have concluded over this city because the city has the least amount of data and hence it is the city which , neutral for comparing all the models. Hence we can conclude that for time series data machine learning models still have to work through and compete with statistical models.



\end{document}